\documentclass[letterpaper,twocolumn,superscriptaddress,floatfix]{revtex4}
\usepackage[latin1]{inputenc}
\usepackage{bm}
\usepackage{multirow,amssymb,amsbsy,amsmath}
\usepackage{graphicx}
\usepackage{verbatim}
\makeatletter
\usepackage{pifont}
\makeatother

\begin{document}

\title{Efficient spectral hole-burning and atomic frequency comb storage in Nd$^{3+}$:YLiF$_4$}

\author{Zong-Quan Zhou}
\affiliation{Key Laboratory of Quantum Information, University of
Science and Technology of China, CAS, Hefei, 230026, China}
\author{Jian Wang}
\affiliation{Key Laboratory of Quantum Information, University of
Science and Technology of China, CAS, Hefei, 230026, China}
\author{Chuan-Feng Li$\footnote{email:cfli@ustc.edu.cn}$}
\affiliation{Key Laboratory of Quantum Information, University of
Science and Technology of China, CAS, Hefei, 230026, China}
\author{Guang-Can Guo}
\affiliation{Key Laboratory of Quantum Information, University of
Science and Technology of China, CAS, Hefei, 230026, China}

\date{\today}

\maketitle
\textbf{
We present spectral hole-burning measurements of the $^{4}I_{9/2}\rightarrow{ }^4F_{3/2}$ transition in Nd$^{3+}$:YLiF$_4$. The isotope shifts of Nd$^{3+}$ can be directly resolved in the optical absorption spectrum. We report atomic frequency comb storage with an echo efficiency of up to 35\% and a memory bandwidth of 60 MHz in this material. The interesting properties show the potential of this material for use in both quantum and classical information processing.}

Due to their long optical and spin coherence times, rare-earth (RE) ion doped solids have shown excellent properties as candidate solid state quantum memories \cite{Tittle09,review10,review101}. Several protocols have been proposed for storing single photons in RE doped solids, for example, electromagnetically induced transparency \cite{sl,time05}, controlled reversible inhomogeneous broadening \cite{crib01,crib06,crib08,crib10,crib062,eff10} and atomic frequency comb (AFC) \cite{AFC09,AFC08}. Among these protocols, AFC has shown better performance in large bandwidth \cite{bw3} and multi-mode capacities \cite{mode,bw1,bw2}, which are especially important for quantum repeater applications. The AFC protocol requires a pumping procedure to obtain an absorption profile with a series of periodic and narrow absorbing peaks separated by $\Delta$ in the frequency domain. The input photons are then collectively absorbed and diffracted by the atomic frequency grating.  Due to the periodic structure of AFCs, an atomic rephasing occurs after a time $\tau_s=1/\Delta$. The photon is re-emitted as a result of a collective interference between all of the atoms that are in phase \cite{AFC09}. The collective optical excitation can be transferred into a long-lived ground state to achieve a longer storage time and an on-demand readout \cite{AFCspin,spin2012,spin2013}.

Currently, AFC storage has been successfully implemented in several RE doped solids, including Nd$^{3+}$:YVO$_4$ \cite{AFC08,polarization}, Nd$^{3+}$:Y$_2$SiO$_5$ \cite{bw1,timebin,entangle}, Tm$^{3+}$:YAG \cite{bw2,Tm1,Tm2,Tm3,Tm4}, Pr$^{3+}$:Y$_2$SiO$_5$ \cite{AFCspin,Pr1,AFCeff,Pr2,Pr3}, Eu$^{3+}$:Y$_2$SiO$_5$ \cite{spin2012,spin20131}, Er$^{3+}$:Y$_2$SiO$_5$ \cite{Er} and Tm$^{3+}$:LiNbO$_3$ \cite{bw3,TmLi}. Nd$^{3+}$ ions are of particular interests because they can provide large memory bandwidth and their working wavelengths are around 870 nm, a wavelength region where diode lasers and efficient single photon detectors are available.
Here we demonstrate efficient optical pumping in a Nd$^{3+}$:YLiF$_4$ crystal. A spin polarization of better than 99.5\% is achieved. We identify the experimental conditions neccessary to achieve efficient AFC storage in this material, performance of which is comparable with that of other RE doped solids.

The experimental sample is a Nd$^{3+}$:YLiF$_4$ crystal (doping level: 100 ppm). The length of the crystal is 2 mm along the a-axis. The $^{4}I_{9/2}\rightarrow{ }^4F_{3/2}$ transition ($\sim$867 nm) in Nd$^{3+}$:YLiF$_4$ shows strong absorption for V-polarized light and no detectable absorption for H-polarized light. Here H (V) denotes horizontal (vertical), which is defined to be parallel (perpendicular) to the crystal's c-axis. The experimental setup is similar to those used in our previous work \cite{polarization,setup}. To carry out spectral hole-burning measurements, two beams of light are independently controlled by two acousto-optic modulators (AOM) in double-pass configurations. The pump light and probe light are then combined and collected with a single mode fiber before being sent into the sample. The transmitted probe signal is detected with a 150 MHz detector. The linearity of the detector in the detecting power range is experimentally verified using an optical power meter.

%%%%%%%%%%%% FIGURE 1 %%%%%%%%%%%%%%%%%%%%%%%%%%%%%%%%%%
\begin{figure}[tb]
\centering
\includegraphics[width=0.5\textwidth]{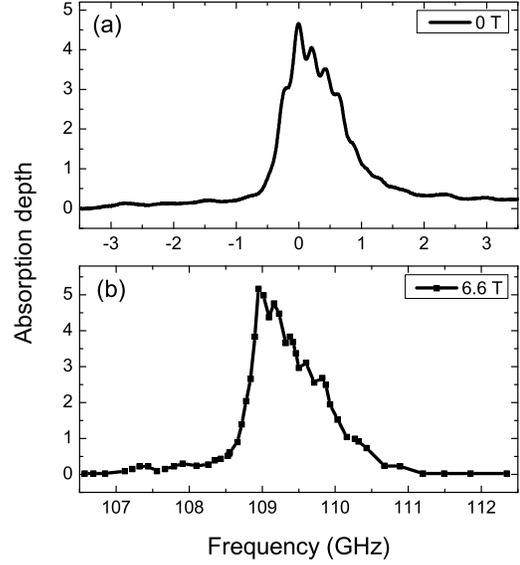}

\caption{\label{Fig:1} The absorption of V-polarized light by the sample at a temperature of 1.5 K and with a magnetic field of 0 T (a) and 6.6 T (b). The optical frequency is set as frequency shift to the absorption peak (345829.65 GHz) of the $^{142}$Nd$^{3+}$ without a magnetic field. The even mass number isotopes can be clearly resolved, with an isotope shift of approximately 110 MHz/unit mass.}
\end{figure}
%%%%%%%%%%%%%%%%%%%%%%%%%%%%%%%%%%%%%%%%%%%%%%%%%%%%%%%%

We first measure the absorption spectrum of our sample. The sample is placed in a cryostat at a temperature of 1.5 K and with a superconducting magnetic field of up to 7 T. The absorption of V-polarized light by the sample for the case of zero magnetic field is shown in Fig.1 (a). The frequency of the Ti:sapphire laser is scanned over 7 GHz and monitored with a wavelength meter. The strongest absorption peak is resulted from the isotope $^{142}$Nd. Compared with previous measurements on the low doped Y$^7$LiF$_4$ crystal \cite{isotope}, we can identify the four peaks at the right of the strongest peak as other even isotopes. The relative intensities of the lines correspond approximately with the percentage natural abundance: 27.1\% $^{142}$Nd, 23.9\% $^{144}$Nd, 17.8\% $^{146}$Nd, 5.7\% $^{148}$Nd and 5.6\% $^{150}$Nd. The correspondence is not exact because some of the odd isotope (natural abundance: 12.2\% $^{143}$Nd, 8.3\% $^{145}$Nd) hyperfine lines overlap with the even isotope lines \cite{isotope}. Due to high doping level of Nd and the naturally occurring Li used, the isotope lines are less clearly resolved compared to that obtained in \cite{isotope}.

The absorption spectrum with a magnetic field of 6.6 T along the crystal's c-axis is shown in Fig.1 (b). To avoid spectral-hole burning, we send short laser pulses to probe the absorption. The inhomogeneous broadening becomes narrower for a higher field and so the five different even isotopes can be better resolved in this case than they can with zero field. We have measured the transition frequency under different magnetic fields. The transition frequency of the even istopes shows a magnetic field dependence of 16.45 $\pm$ 0.09 GHz/T.
%%%%%%%%%%%% FIGURE 2 %%%%%%%%%%%%%%%%%%%%%%%%%%%%%%%%%%
\begin{figure}[tb]
\centering
\includegraphics[width=0.5\textwidth]{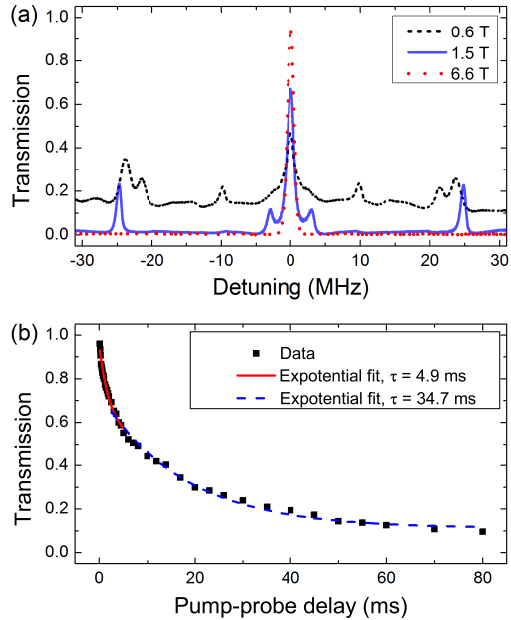}
\caption{\label{Fig:2} (a). The transmission profiles with the pump light set to burn a spectral hole at the central frequency. The black dashed, blue solid and red dotted lines show measurements with $\textmd{T}_\textmd{d}$=0.2 ms and a magnetic field of 0.6 T, 1.5 T and 6.6 T, respectively. (b). The decay of transmission of the spectral hole as a function of pump and probe delay $\textmd{T}_\textmd{d}$. The spectral hole shows two rates of decay.}
\end{figure}
%%%%%%%%%%%%%%%%%%%%%%%%%%%%%%%%%%%%%%%%%%%%%%%%%%%%%%%%

The background of the technique of spectral hole-burning spectroscopy can be found in Ref. \cite{SHB,Gisin10}. To burn a spectral hole, the pump light is set at the central frequency and the pump sequence takes a time of 10 ms. After a waiting time of $\textmd{T}_\textmd{d}$, the probe light is sent into the sample and probes the transmission profiles. The probe light is swept over 60 MHz by the AOM. The measured spectral hole-burning spectra are shown in Fig. 2 (a) for setups with a magnetic field of 0.6 T, 1.5 T and 6.6 T. The nonuniform frequency response of the probe AOM has been corrected. With a magnetic field applied, the absorption band begins to split into two groups of lines and thus the absorption is weaker for a small magnetic field than for zero field. The inhomogeneous broadening continues to narrow with growing magnetic field, so the peak absorption with high magnetic field is comparable with that of zero field. The minimum absorption depth of the spectral hole is 0.05 with a magnetic field of 6.6 T. Considering that the total population is initially equally distributed in the two spin states, only $\sim$0.5\% of the total population is left in the initial state. This corresponds to a spin polarization of 99.5\%, which is significantly better than that obtained in Nd$^{3+}$:YVO$_4$ \cite{Gisin10}. One can observe several side holes in the hole-burning spectra for a magnetic fields of 0.6 T and 1.5 T. To achieve efficient AFC storage, there should be no side holes or anti-holes within the memory bandwidth. A high field is necessary to move these side holes out of the storage band. With a magnetic field of 6.6 T, no side holes or anti-holes can be observed within the 60 MHz absorption band.

To characterize the lifetime of the population trapping, we measure the dynamics of the central spectral hole. The hole transmission is probed with readout pulses with delays $\textmd{T}_\textmd{d}$ of up to 80 ms. The measured transmissions are shown with black squares in Fig. 2 (b). From Fig. 2 (b), there appear to be two rates of decay in this material, measured by fitting $e^{-2\textmd{T}_\textmd{d}/\tau}$. There is an initial decay with a decay constant of 4.9 ms, followed by a secondary decay with a decay constant of 34.7 ms. The reason for the two decay rates is unknown. One possible explanation could be different isotopes decaying at different rates.

Because very 'clean' spectral hole-burning spectrum is obtained with a magnetic field of 6.6 T, a high contrast AFC should be easily achieved at the same conditions. The pump AOM is then programmed to produce a periodical structure in the frequency domain (see methods).
Fig. 3(a) shows an example of coherent pulses with durations of approximately 60 ns that are collectively mapped onto the sample. A strong echo is emitted after a preprogrammed storage time of 143 ns. The measured storage and retrieval efficiency is approximately 35\%. The storage efficiency is better than that obtained in Nd$^{3+}$ doped YVO$_4$ \cite{polarization,AFC08} and Y$_2$SiO$_5$ \cite{bw1,entangle,timebin} crystals.

%%%%%%%%%%%% FIGURE 3 %%%%%%%%%%%%%%%%%%%%%%%%%%%%%%%%%%
\begin{figure}[tb]
\centering
\includegraphics[width=0.5\textwidth]{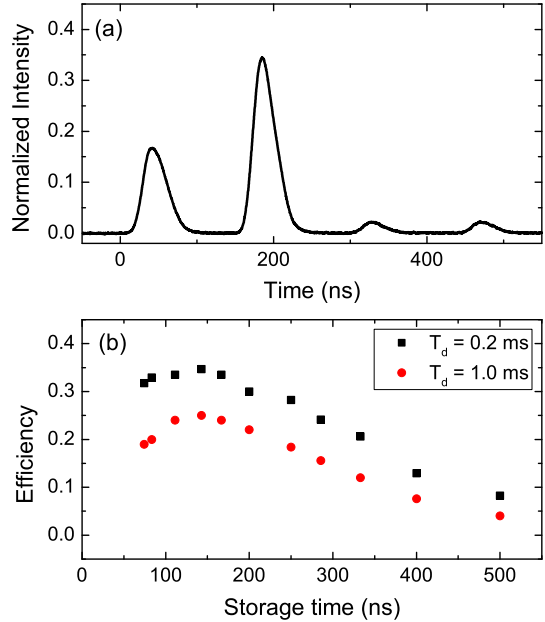}
\caption{\label{Fig:3} (a). With an AFC prepared with a periodicity of 7 MHz,
the optical pulses are collectively re-emitted after a 143 ns storage time in the sample. (b). The AFC storage efficiency as a function of storage time. The black squares and red dots show measurements with $\textmd{T}_\textmd{d}$ of 0.2 ms and 1.0 ms, respectively.}
\end{figure}
%%%%%%%%%%%%%%%%%%%%%%%%%%%%%%%%%%%%%%%%%%%%%%%%%%%%%%%%
The storage efficiencies measured at different storage times are shown in Fig.3 (b), with black squares and red dots representing measurements with $\textmd{T}_\textmd{d}$ of 0.2 ms and 1.0 ms, respectively. Because of the fast initial decay of population trapping, the storage efficiency is significantly higher for $\textmd{T}_\textmd{d}$ of 0.2 ms. For shorter storage time, the efficiency shows a small drop because the long input pulses have limited bandwidth. The input pulses can cover less AFC teeth for shorter storage time.

We note that, unlike the experiments performed in Nd$^{3+}$:YVO$_4$ crystals, the pump power consumptions for AFC preparations are much higher for Nd$^{3+}$:YLiF$_4$ crystals. The maximum pump power we can produce is approximately 80 mW, which is limited by the damage threshold of the AOM and the single mode fiber coupling loss. For limited pump power, the maximum storage efficiency is achieved by shifting the laser frequency away from the absorption peaks. As shown by the red solid line in Fig. 4, the initial absorption depth without pump light is approximately 2.6. After the pumping procedures, an AFC with periodicity of 7 MHz is engineered in the sample. As shown by the black dotted line in Fig. 4, the background absorption $d_0$ is approximately 0.05 and the peak absorption $d$ is approximately 5.5. The storage efficiency can be estimated by $\eta\approx e^{-d/F}(d/F)^2 e^{-7/F^2}e^{-d_0}$, where $F\equiv\Delta/\gamma$ is the fineness of the comb and $\gamma$ is the linewidth of the teeth \cite{AFC09}. The experimentally measured efficiency is slightly higher than expectated ($\sim$30\%).

%%%%%%%%%%%% FIGURE 4 %%%%%%%%%%%%%%%%%%%%%%%%%%%%%%%%%%
\begin{figure}[tb]
\centering
\includegraphics[width=0.5\textwidth]{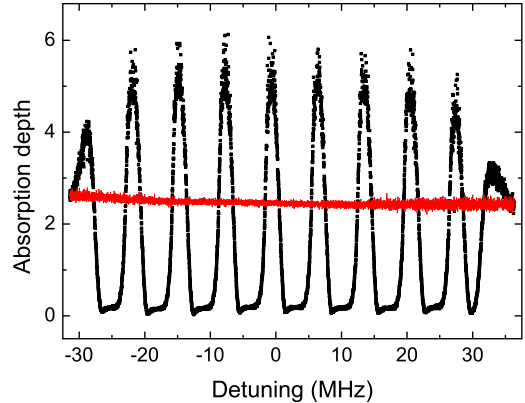}
\caption{\label{Fig:4} Tailored absorption band. The red solid line shows the initial absorption without any pump procedures. The black dotted line shows an AFC prepared with a periodicity of 7 MHz. The background absorption $d_0$ is approximately 0.05, while the absorption peaks show a strong absorption $d\sim5.5$, which is larger than the initial absorption.}
\end{figure}
%%%%%%%%%%%%%%%%%%%%%%%%%%%%%%%%%%%%%%%%%%%%%%%%%%%%%%%%

An interesting phenomenon is that the frequency components without pump light show a much stronger absorption than the initial absorption. A similar phenomenon has been previously observed in other RE-doped solids \cite{Tittel}. However, in \cite{Tittel}, the absorption rose only a small amount after spectral tailoring. The population redistribution behavior could possibly be caused by anti-holes of some hyperfine structures that are not clearly resolved in Fig. 2 (a). The contrast ratio between absorption peaks and background absorption $d/d_0$ is approximately 100:1. The absorption is optically controlled by the pump light, so we suggest that this material may be used as an optically controlled optical shutter. The efficient AFC structure can be engineered on a weakly absorbing sample, thus, shorter crystal length can be tolerated. This is important for future applications aimed at making integrated devices.

The present work demonstrated a two-level AFC storage, which is only a programmable delay line. A complete AFC storage requires a third ground level to achieve long storage time and on-demand readout. The $^{143}$Nd$^{3+}$ has a nuclear spin of 7/2 and exhibits hyperfine splittings under zero magnetic field, which should be able to provide the required energy levels and optical transitions. The isotope-dependent inhomogeneous absorption opens the possibility to frequency selectively address the $^{143}$Nd$^{3+}$ in a sample with Nd$^{3+}$ ions of natural abundance. Further experiments with high doping crystals would be required to reach this goal. To avoid the influence of other isotopes, isotopically enriched $^{143}$Nd$^{3+}$:YLiF$_4$ would be a better choice for complete AFC storage.
Until now, complete AFC storage has only been demonstrated in Pr$^{3+}$ and Eu$^{3+}$ doped solids \cite{AFCspin,spin2012,spin2013}, whose working wavelengths are around 600 nm, a range that is not covered by flexible semiconductor lasers. Their memory bandwidth are fundamentally limited by the small hyperfine splittings of Pr$^{3+}$ and Eu$^{3+}$, which are several MHz to tens of MHz. According to the spectroscopy of Nd$^{3+}$ isotopes obtained in \cite{isotope}, the hyperfine splitting of $^{143}$Nd$^{3+}$ are hundreds of MHz, which provide potentially large bandwidth of quantum memory. Further study on $^{143}$Nd$^{3+}$:YLiF$_4$ will be required to identify these transitions for implementation of the spin-wave storage in this material. The complete AFC storage implemented in Nd$^{3+}$ doped solids is more appealing because of the suitable working wavelength and potentially large bandwidth.

We have presented spectral hole-burning measurements in Nd$^{3+}$:YLiF$_4$. With a temperature of 1.5 K and a magnetic field of 6.6 T, AFC storage is achieved in this material with a storage efficiency of up to 35\% and with memory bandwidth of 60 MHz. A population redistribution behavior in the frequency domain is observed in the hole-burning spectrum. These properties make this material an interesting candidate for various information processing protocols in both quantum and classical networks.

\section*{Methods}
\textbf{AFC preparation.}
Tailoring the absorption profile requires simultaneously sweeping the pump light frequency and modulating its intensity. In our experiment, the pump light's frequency is swept over 60 MHz in 240 $\mu$s, and each frequency step has been assigned a specific amplitude to produce the desired structure. This procedure ensures a spectral tailoring resolution of approximately 0.5 MHz. The pump sequences are continuously repeated over 10 ms to achieve an optimal absorption structure. The pump light power and the duty ratio in each cycle are carefully adjusted to take into account the power broadening effect. The AOMs are controlled by programmable RF drivers using a Universal Serial Bus interface. Control of the pump light can be fully realized through computer interfaces. The pump and probe timing are controlled by arbitrary function generators (Tektronix, AFG3252).

%%%%%%%%%%%%%%%%%%%%%%%%%%%%%%%%%%%%%%%%%%%%%%%%%%%%%%%%

{\bf  Acknowledgments}

This work was supported by the CAS, the National Basic Research Program (2011CB921200), National Natural Science Foundation of China (Grant No. 60921091 and No. 11274289) and the Fundamental Research Funds for the Central Universities (wk2030380004 and wk2470000011).

{\bf Author Contributions}

C-F.L. and Z-Q.Z. planned and designed the experiments. Z-Q. Z. and J. W. implemented the experiments. Z-Q.Z., C-F.L. and G-C.G. wrote the paper. C-F.L. supervised the project.

\section*{Additional Information}
\subsection*{Competing Financial Interests}
The authors declare no competing financial interests.

\clearpage
\end{document}